%% file: CommentsProakisAnalysisQuadraticforms.tex
\documentclass[twocolumn,twoside]{IEEEtran}           

\usepackage{amsmath,amssymb,amsfonts}
\usepackage[final]{graphicx} 

\usepackage{psfrag}

\def \H {{\mathcal{H}}}
\def \E {{\text{E}}}

\begin{document}

\title{Comments on Proakis' Analysis of the Characteristic Function of Complex Gaussian Quadratic Forms} 

\author{Unai Fern\'andez-Plazaola, Eduardo Martos-Naya, Jos\'e F. Paris \mbox{and Jos\'e T. Entrambasaguas}}

\markboth{Technical Report. Dpto. Ingenier\'ia de Comunicaciones.}
{Fern\'andez, Martos and Paris: Comments on Proakis ...}

\maketitle

\begin{abstract}
An analysis of the characteristic function of Gaussian quadratic
forms is presented in \cite{Proakis68} to study the performance of
multichannel communication systems. This technical report reviews
this analysis, obtaining alternative expressions to original ones
in compact matrix format.
\end{abstract}


\section{Introduction}

\PARstart{S}{tatistics} of complex Gaussian quadratic forms are a
powerful mathematical tool when dealing with digital
communications problems. Currently, they are widely used to
analyze the performance of digital modulations over fading
channels \cite{Proakis01}-\cite{McCloud02}. In \cite{Proakis68} is
presented an analysis of one statistic of complex Gaussian
quadratic forms which is a very interesting and useful tool to
calculate error rates. This analysis, based on the characteristic
function of Gaussian quadratic forms, provides a set of
expressions which is applicable to a good deal of digital
communications problems. To provide a few examples, these
expressions have been used to obtain closed-form bit error rate
(BER) results in fading channels for multichannel binary signals
\cite{Proakis01}-\cite{Sudhanshu03}, quadrature-amplitude
modulation (QAM) \cite{Cao05},
orthogonal-frequency-division-multiplexing (OFDM) \cite{Jun00},
and non-orthogonal multi-pulse modulation (NMM) \cite{McCloud02}.
In addition, the results presented in \cite{Proakis68} are also
used to analyze other statistics of complex Gaussian quadratic
forms in \cite{Biyari93}.

In this document is performed a revision to the analysis of the
characteristic function of complex gaussian quadratic forms
presented in \cite{Proakis68}. Alternative expressions in compact
matrix format are derived and little mistakes in the original ones
has been detected. Up until now, research based on the original
expressions has yielded valid results because it dealt with cases
where the original and corrected expressions give the same
results. However, future uses of the original expressions could
lead to wrong results. After deriving alternative expressions, a
simple example is presented to confirm the validity of the
corrected expressions.

The remainder of this document is organized as follows: Section
\ref{general} states the general problem dealt with herein;
Section \ref{expressions} describes the alternative analysis which
gives the new expressions and detects the mistakes in the original
expressions; Section \ref{example} presents the aforementioned
example which confirms the validity of the new ones; and
conclusions are provided in section \ref{conclusions}.



\section{General problem statement}
\label{general}

The decision variable at the detector of several communication systems, e.g., those employing
multichannel binary signals \cite{Proakis68}, can be expressed as a sum of quadratic forms

\begin{equation}
\begin{split}
D & \triangleq \sum\limits_{k = 1}^L {d_k } = \sum\limits_{k = 1}^L {\mathbf{z}}_k^{\H}
{\mathbf{Qz}}_k \\
&= \sum\limits_{k = 1}^L A\left| {X_k } \right|^2  + B\left| {Y_k } \right|^2 + CX_k Y_k^*  +
C^* X_k^* Y, \label{eq:general:1}
\end{split}
\end{equation}
with
\begin{equation}
 {\mathbf{z}}_k  \triangleq \left[ {\begin{array}{*{20}c}
   {X_k }  \\
   {Y_k }  \\
 \end{array} } \right]
 \quad , \quad
 {\mathbf{Q}} \triangleq \left[ {\begin{array}{*{20}c}
   A & {C^* }  \\
   C & B  \\
 \end{array} } \right]  \quad , \quad d_k \triangleq {\mathbf{z}}_k^{\H}
{\mathbf{Qz}}_k, \label{eq:general:2}
\end{equation}
where $\mathbf{Q}$ is a Hermitian matrix and $X_k$, $Y_k$ complex circularly-symmetric
Gaussian random variables. The set of vectors $\{{\mathbf{z}}_k\}_{k=1,...,L}$ are mutually
statistically independent with a different mean ${\mathbf{m}}_k$ but identical non-singular
covariance matrix $\mathbf{R}$, respectively defined as
\begin{equation}
{\mathbf{m}}_k  \triangleq \left[ {\begin{array}{*{20}c}
   {\bar X_k }  \\
   {\bar Y_k }  \\
 \end{array} } \right]  \quad , \quad
  {\mathbf{R}} \triangleq
 \E\left\{ {({\mathbf{z}}_k  - {\mathbf{m}}_k )({\mathbf{z}}_k  -
{\mathbf{m}}_k )^\H } \right\}.
 \label{eq:general:3}
\end{equation}

The constants $A$, $B$ and $C$ must be appropriately identified with the specific parameters
of the problem, so that $D$ characterizes the decision variable at the output of the detector
to further calculate the probability of error as $\Pr\{D<0\}$. This calculation leads to
non-trivial results when $\mathbf{Q}$ is indefinite, i.e., when $\left| C \right|^2 - AB
> 0$, because otherwise the probability $\Pr\{D<0\}$ will be either 0 or 1. Note that
indefiniteness for $2 \times 2$ matrices implies non-singularity.

An expression for the probability \mbox{$\Pr\{D<0\}$} is derived in \cite{Proakis68} using the
characteristic function of $D$, denoted as $\phi _D (\upsilon)$. In this derivation, mutual
independence of the terms $d_k$ in (\ref{eq:general:1}) is used, thus, the function $\phi _D
(\upsilon)$ can be expressed as the product of the characteristic functions $\phi_k$ of each
summand, i.e.,
\begin{equation}
 \phi _D (\upsilon) \triangleq  \E\left\{ {e^{j \upsilon D} } \right\} = \prod\limits_{k = 1}^L {\phi _k (\upsilon)}
\label{eq:general:4},
\end{equation}
with
\begin{equation}
\phi_k (\upsilon) = \frac{{v_1 v_2 }} {{\left( {\upsilon + jv_1 } \right)\left( {\upsilon -
jv_2 } \right)}} \exp \left( {\frac{{v_1 v_2 \left( { - \upsilon ^2 \alpha _{1k} + j\upsilon
\alpha _{2k} } \right)}} {{\left( {\upsilon  + jv_1 } \right)\left( {\upsilon  - jv_2 }
\right)}}} \right) \label{eq:general:5}.
\end{equation}
The first two columns of table \ref{table:1} lists the parameters $v_1$, $v_2$, $\alpha_{1k}$,
$\alpha_{2k}$, as well as other intermediate definitions and the expression of the probability
\mbox{$\Pr\{D<0\}$} derived in \cite{Proakis68}. In the table, $Q_1$ and $I_n$ are the
first-order Marcum Q-function and the modified Bessel function of the first kind,
respectively.

\begin{table*}[tbp]

\input{table1.tex}

\input{table2.tex}
\end{table*}

\section{Alternative analysis}
\label{expressions}

An alternative formulation of the characteristic function $\phi_k$ is used in this section to
obtain new expressions for the probability \mbox{$\Pr\{D<0\}$}.

In \cite{turin97}, Turin deducts the following expression for the characteristic function of a
quadratic form of $n$ variables $\mathbf{z}^\H \mathbf{Qz}$
\begin{equation}
\phi (\upsilon) = \frac{{\exp \left( { - {\mathbf{m}}^\H {\mathbf{R}}^{ - 1} \left[
{{\mathbf{I}} - ({\mathbf{I}} - j \upsilon {\mathbf{RQ}})^{ - 1} } \right]{\mathbf{m}}}
\right)}} {{\left| {{\mathbf{I}} - j\upsilon {\mathbf{RQ}}} \right|}},
 \label{eq:analysis:1}
\end{equation}
where $\mathbf{Q}$ is an $n \times n$ Hermitian matrix and $\mathbf{z}$ a complex Gaussian
vector of $n$ dimensions with mean $\mathbf{m}$ and covariance $\mathbf{R}$. The matrix
$\mathbf{R}$ is assumed to be non-singular, as it is a covariance matrix, is therefore
positive definite.

Expression (\ref{eq:analysis:1}) only makes sense if the determinant $\left| {{\mathbf{I}} -
j\upsilon {\mathbf{RQ}}} \right|$ is non-zero. Proof of this fact is easy to find if the
determinant is expressed as $\prod\nolimits_{i = 1}^n {\left( {1 - j\upsilon \delta _i }
\right)}$ \cite{turin97}, where $\{\delta_i\}$ represents the eigenvalues of $\mathbf{RQ}$,
which must be real so that the determinant will be non-zero. These eigenvalues, by definition,
satisfy $\left| \mathbf{RQ} - \delta_i \mathbf{I} \right| = 0$, which can be rearranged as
\mbox{$\left|\mathbf{R}^{1/2} \mathbf{Q} \mathbf{R}^{1/2} - \delta_i \mathbf{I} \right| = 0$},
where $\mathbf{R}^{1/2}$ is the matrix square root of $\mathbf{R}$, which is also a Hermitian
positive definite matrix. Hence, as matrix $\mathbf{R}^{1/2} \mathbf{Q} \mathbf{R}^{1/2}$ is
Hermitian, the eigenvalues $\{\delta_i\}$ are real and therefore $\left| {{\mathbf{I}} -
j\upsilon {\mathbf{RQ}}} \right|$ is non-zero.

The characteristic function $\phi$ in (\ref{eq:analysis:1}) can be rearranged as
\begin{equation}
\begin{split}
\phi(\upsilon)& = \frac{{\exp ( - \mathbf{m}^\H \mathbf{R}^{ - 1}
\left[ {(\mathbf{I} - j\upsilon \mathbf{RQ} -
\mathbf{I})(\mathbf{I} - j\upsilon \mathbf{RQ})^{ - 1} }
\right]{\mathbf{m}})}}
{{\left| {\mathbf{I} - j\upsilon \mathbf{RQ}} \right|}} \\
&= \frac{{\exp \left( {j\upsilon {\mathbf{m}}^\H {\mathbf{Q}}({\mathbf{I}} - j \upsilon
{\mathbf{RQ}})^{ - 1} {\mathbf{m}}} \right)}} {{\left| {{\mathbf{I}} - j
\upsilon{\mathbf{RQ}}} \right|}}.
 \label{eq:analysis:5}
\end{split}
\end{equation}
This expression, particularized to $2 \times 2$ quadratic forms, is applicable to the variable
$d_k$ of (\ref{eq:general:2}) to obtain $\phi_k$. Otherwise, it is easy to show that for
$2\times2$ non-singular matrices ${\mathbf{M}}_1$ and ${\mathbf{M}}_2$, whose sum is also
non-singular, the following property is fulfilled
\begin{equation}
({\mathbf{M}}_1 + {\mathbf{M}}_2)^{ - 1}  = \frac{{\left| {\mathbf{M}}_1
\right|{\mathbf{M}}_1^{ - 1} + \left|{\mathbf{M}}_2 \right|{\mathbf{M}}_2^{ - 1} }} {\left|
{{\mathbf{M}}_1 + {\mathbf{M}}_2} \right|}.
 \label{eq:analysis:7}
\end{equation}
Using this property in expression (\ref{eq:analysis:5}) particularized to variable $d_k$, and
remembering that $\mathbf{Q}$ is non-singular, gives 
\begin{equation}
\begin{split}
\phi _k (\upsilon) &= \frac{1} {{\left| {{\mathbf{I}} - j \upsilon {\mathbf{RQ}}} \right|}}\\
& \cdot \exp \left( {\frac{{j \upsilon {\mathbf{m}}_k^\H {\mathbf{Qm}}_k + \upsilon^2 \left|
{\mathbf{RQ}} \right|{\mathbf{m}}_k^\H {\mathbf{R}}^{ - 1} {\mathbf{m}}_k}} {{\left|
{{\mathbf{I}} - j \upsilon {\mathbf{RQ}}} \right|}}} \right),
 \label{eq:analysis:9}
\end{split}
\end{equation}
or, if $\delta_1$ and $\delta_2$ are the eigenvalues of matrix $\mathbf{RQ}$,
\begin{equation}
\begin{split}
\phi_{k} (\upsilon) &= \frac{1} {{(1 - j \upsilon \delta_1 )(1 - j \upsilon \delta_2 )}}\\
& \cdot \exp \left({\frac{{j \upsilon {\mathbf{m}}_k^\H {\mathbf{Qm}}_k + \upsilon^2 \delta_1
\delta_2{\mathbf{m}}_k^\H {\mathbf{R}}^{ - 1} {\mathbf{m}}_k}} {{(1 - j \upsilon \delta_1 )(1
- j \upsilon \delta_2 )}}} \right).
 \label{eq:analysis:10}
\end{split}
\end{equation}

It is easy to show that the eigenvalues of matrix $\mathbf{RQ}$ can be defined as
\begin{equation}
\left\{ {\delta _i } \right\}_{i = 1,2}  = \dfrac {\text{tr}({\mathbf{RQ}}) \pm \sqrt
{\text{tr}({\mathbf{RQ}})^2 - 4 \; \left|{\mathbf{RQ}} \right|}} {2}, \label{eq:analysis:11}
\end{equation}
where $\text{tr}(\cdot)$ represents the trace. The positive sign of (\ref{eq:analysis:11}) has
been chosen for eigenvalue $\delta_1$ and the negative sign for eigenvalue $\delta_2$. In
order to obtain non-trivial results for \mbox{$\Pr\{D<0\}$}, as mentioned in the previous
section, matrix $\mathbf{Q}$ must be indefinite. Hence, as matrix $\mathbf{R}$ is positive
definite, matrix $\mathbf{RQ}$ is indefinite. So, $\delta_1$ takes a strictly positive value
and $\delta_2$ a strictly negative value.

Comparing (\ref{eq:analysis:10}) and (\ref{eq:general:5}) it is possible to find the following
alternative definitions for parameters $v_1$, $v_2$, $\alpha_{1k}$, $\alpha_{2k}$,
\begin{equation}
\left\{ {\begin{array}{*{20}c}
   {v_1 = 1 / \delta_1} \hfill \\
   {v_2 = - 1 / \delta_2} \hfill \\
   {\alpha _{1k}  = -\delta_1 \delta_2 {\mathbf{m}}_k^\H {\mathbf{R}}^{ - 1} {\mathbf{m}}_k} \hfill \\
   {\alpha _{2k}  = {\mathbf{m}}_k^\H {\mathbf{Qm}}_k} \hfill \\
 \end{array} } \right.
\label{eq:analysis:12}
\end{equation}

These definitions are now in a compact matrix format and second
they make it easier to find little mistakes in original
expressions. This mistakes are clarified in table \ref{table:1},
where the original parameters are listed in one column and the
parameters with mistakes are corrected in the other column. It is
important to highlight that the detected mistakes only affect
parameters $w$ and $\alpha_{1k}$ insofar as they are related to
constant $C$. Thus, the original expressions are only wrong when
the constant $C$ is complex.

New expressions of the probability $\Pr\{D<0\}$ can be used taking
into account (\ref{eq:analysis:12}). These expressions only needs
the two eigenvalues, $\delta_1$ and $\delta_2$, and parameters $a$
and $b$, and they are summarized in table \ref{table:2}.

In most cases where the original expressions have been used up
until now, the constant $C$ was real, so the results obtained were
valid. For example, in chapter 12 of \cite{Proakis01}, in the
context of multichannel digital communication with binary
signaling, two types of processing at the receiver are considered,
coherent and non-coherent detection. For the coherent detector the
constant $C$ is equal to 1/2 and for the non-coherent detector the
constant $C$ is equal to 0. Thus, in both cases the constant $C$
is real and the mistakes in the expressions do not affect to the
results. Another example is presented in \cite{Cao05}, where
Gray-code 16-QAM constellation is considered. In this case, the
decision variables for the in-phase and quadrature components are
defined with constant $C$ equal to $1/2$ and $-j/2$, respectively.
Despite the fact that constant $C$ is complex for the quadrature
component, the symmetry of the problem was taken into account in
\cite{Cao05}, making it possible to perform the calculations in
another way in which $C$ was real. Therefore, the results obtained
in \cite{Cao05} are valid but would be wrong if the problem were
solved using the quadrature component variable decision ($C$
complex). A simple example where $C$ is complex is shown in the
next section. Again, the example serves to show the mistakes in
the original expressions, revealing how they lead to incorrect
results and showing how the new expressions yield correct results.

\section{Example}
\label{example}

In order to justify the implications of the mistakes in the
original expressions, a simple example where the parameter $C$ is
complex is presented in this section. The example chosen has the
following decision variable 
\begin{equation}
\begin{split}
D = & \exp (j \frac {\pi}{4} ) X_1 Y_1^*  + \exp (-j \frac {\pi}{4} ) X_1^* Y_1  \\
= & 2\operatorname{Re} \left\{ {\exp (j \frac {\pi}{4} )X_1 Y_1^* } \right\},
 \label{eq:example:1}
\end{split}
\end{equation}
so, $L=1$ and the constants are identified as $A=0$, $B=0$, and $C=\exp (j \frac {\pi}{4})$.
The complex Gaussian variables $X_1$ and $Y_1$ are chosen to be independent, with mean and
variance \mbox{$\bar X_1 = \exp (j \frac {\pi}{4})$}, $\bar Y_1 = 1$, and $\E\left\{
{\left|X_1-\bar X_1\right|^2 } \right\} = \E\left\{ {\left|Y_1-\bar Y_1\right|^2 } \right\} =
1$, respectively.

\input{table3.tex}

Taking into account the definitions of table \ref{table:1}, all parameters are calculated in
both ways, with original definitions and with corrected definitions, as presented in
\mbox{table \ref{table:3}}. Note that the value of $\Pr\{D<0\}$ calculated with the corrected
expressions is exactly $1/2$, for which \mbox{\cite[eq.(4.53)]{Simon05}} has been used. As
shown in table \ref{table:3}, the two types of definitions give different values for the last
three parameters and the final probability. The example is chosen so that the final
probability can be found easily in another way as follows.

The probability $\Pr\{D<0\}$ can be directly calculated from the definition
(\ref{eq:example:1}) of $D$, as
\begin{equation}
\begin{split}
 \Pr\{D<0\} =&  \Pr \left\{ {2\left| {X_1 } \right|\left| {Y_1 } \right|\cos \left( {\frac{\pi }
{4} + \measuredangle X_1  - \measuredangle Y_1 } \right) < 0} \right\}  \\
=& \Pr \left\{ {\sin \left( {\measuredangle Y_1 - Z} \right) < 0} \right\} \\
=& \Pr \left\{ {\sin \left( {U} \right) < 0} \right\},
\end{split}
 \label{eq:example:4}
\end{equation}
with $Z = \measuredangle X_1  - \pi / 4$ and $U = \measuredangle Y_1 - Z$.

Any non-zero mean complex Gaussian variable $V=\rho \exp (j \theta)$ with mean $\bar V =
\Gamma \exp (j \Theta)$ and standard deviation $\sigma$ has the following joint probability
density function (pdf) \cite{lindsey64}
\begin{equation}
p_{\rho,\theta}(\rho,\theta ) = \frac{\rho} {{\pi \sigma ^2 }}\exp \left( { - \frac{{\rho^2 +
\Gamma^2 - 2 \; \Gamma \; \rho \cos (\theta  - \Theta )}} {{\sigma ^2 }}} \right).
\label{eq:example:5}
\end{equation}
It is evident from (\ref{eq:example:5}) that the marginal pdf $p_\theta(\theta )$ has even
symmetry when its mean $\Theta$ is zero.

Note that $X_1$ and $Y_1$ are non-zero mean complex Gaussian variables. Therefore, the
variables  $\measuredangle Y_1$ and $Z$ are statistically distributed as $\theta$, and as they
have zero-mean they have an even pdf. Moreover, the difference $U$ between the independent
variables $\measuredangle Y_1$ and $Z$ also has even symmetry. This implies that the variable
$\sin(U)$ has an even pdf, because it is an odd function applied to a variable with even pdf.
Consequently, the probability $\Pr\{D<0\}$ is exactly $1/2$.

\begin{figure}[t!]
\includegraphics[width=7.75cm]{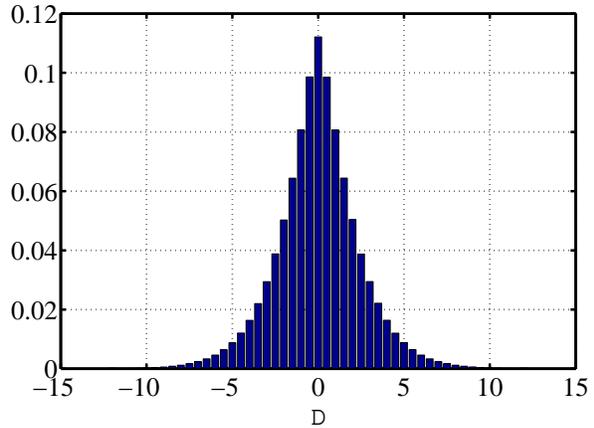}
\caption{Histogram of decision variable $D$} \label{fig:1}
\end{figure}

Figure \ref{fig:1} shows the histogram of the decision variable $D$ obtained by simulations.
Note the even symmetry of the histogram, confirming that $\Pr\{D<0\}=1/2$, the same value
obtained with the corrected expressions.

\section{Conclusions}
\label{conclusions}

In \cite{Proakis68} is presented an analysis of complex Gaussian
quadratic forms which gives some powerful expressions to study
many digital communications systems in terms of the error
probability. This analysis has been revised in this technical
report, obtaining alternative expressions in a compact matrix
format. Little mistakes in the original expressions have been
detected which must be corrected to avoid errors in practical
applications. Finally, a simple example has been presented which
reveals the influence of these mistakes and confirms the validity
of the new expressions.

\end{document}

%% file: table1.tex
\caption{Expressions of the error rate for binary signals}
\begin{center}
\resizebox{!}{7cm}{
\begin{tabular}{c|c|c}
\sc{Parameter} & \sc{Proakis' Definition} & \sc{Correction} \\
 \hline\hline
& &\\
$\mu _{xy}$  &  $ \frac{1} {2}\E\left\{ {\left( {X_k  - \bar X_k } \right)\left( {Y_k  - \bar
Y_k } \right)^* } \right\}$ & \\
\hline
&&\\
$w$       &$\dfrac{{A\mu _{xx}  + B\mu _{yy}  + C \mu _{xy}^*  + C^* \mu _{xy}}} {{4(\mu _{xx}
\mu _{yy}  - \left| {\mu _{xy} } \right|^2 )(\left| C \right|^2  - AB)}}$& $\dfrac{{A \mu
_{xx} + B\mu _{yy}  + C^* \mu _{xy}^*  + C\mu _{xy} }} {{4(\mu _{xx} \mu _{yy} - \left| {\mu
_{xy}
} \right|^2 )(\left| C \right|^2  - AB)}}$\\
\hline
&&\\
 $v_1 $    &$\sqrt {w^2  + \dfrac{1} {{4(\mu _{xx} \mu _{yy}  - \left| {\mu _{xy} } \right|^2
)(\left| C \right|^2  - AB)}}}  - w$ &\\
\hline
&&\\
 $v_2  $    &$\sqrt {w^2  + \dfrac{1} {{4(\mu _{xx} \mu _{yy}  - \left| {\mu _{xy} }
\right|^2
)(\left| C \right|^2  - AB)}}}  + w$ &\\
\hline
&&\\
$\alpha _{1k}$ & $2\left( {\left| C \right|^2  - AB} \right)\left( {\left| {\bar X_k }
\right|^2 \mu _{yy}  + \left| {\bar Y_k } \right|^2 \mu _{xx}  - \bar X_k^* \bar Y_k \mu _{xy}
- \bar X_k \bar Y_k^* \mu _{xy}^* } \right)$\\
\hline
&&\\
$\alpha _{2k}$ & $A\left| {\bar X_k } \right|^2  + B\left| {\bar Y_k } \right|^2  + C \bar
X_k^* \bar Y_k  + C^* \bar X_k \bar Y_k^*$ & $A\left| {\bar X_k } \right|^2 + B\left|
{\bar Y_k }\right|^2  + C^* \bar X_k^* \bar Y_k + C \bar X_k \bar Y_k^*  $ \\
\hline
&&\\
$\alpha _1$ & $ \sum\nolimits_{k = 1}^L {\alpha _{1k} }$ \\
\hline
&&\\
$\alpha _2$ & $ \sum\nolimits_{k = 1}^L {\alpha _{2k} }$ \\
\hline
&&\\
$a$  &  $ \left[ {\dfrac{{2v_1^2 v_2 \left( {\alpha _1 v_2  - \alpha _2 } \right)}} {{\left(
{v_1 + v_2 } \right)^2 }}} \right]^{1/2}$ &\\
\hline
&&\\
$b$  &  $ \left[ {\dfrac{{2v_1 v_2^2 \left( {\alpha _1 v_1  + \alpha _2 } \right)}} {{\left(
{v_1 + v_2 } \right)^2 }}} \right]^{1/2}$ &\\
\hline \hline
&\multicolumn{2}{l}{}\\
$ \Pr\{D<0\}  $& \multicolumn{2}{c}{$\begin{gathered}
  Q_1 (a,b) - I_0 (ab)\exp \left\{ { - \frac{1}
{2}(a^2  + b^2 )} \right\}
   + \frac{{I_0 (ab)\exp \left\{ { - \frac{1}
{2}(a^2  + b^2 )} \right\}}} {{(1 + v_2 /v_1 )^{2L - 1} }}\sum\limits_{k = 0}^{L - 1} {\left(
{\begin{array}{*{20}c}
   {2L - 1}  \\
   k  \\
 \end{array} } \right)\left( {\frac{{v_2 }}
{{v_1 }}} \right)^k }  \hfill \\
   + \frac{{\exp \left\{ { - \frac{1}
{2}(a^2  + b^2 )} \right\}}} {{(1 + v_2 /v_1 )^{2L - 1} }}\sum\limits_{n = 1}^{L - 1} {I_n
(ab)} \sum\limits_{k = 0}^{L - 1 - n} {\left( {\begin{array}{*{20}c}
   {2L - 1}  \\
   k  \\
 \end{array} } \right)}
    \left[ {\left( {\frac{b}
{a}} \right)^n \left( {\frac{{v_2 }} {{v_1 }}} \right)^k  - \left( {\frac{a} {b}} \right)^n
\left( {\frac{{v_2 }}
{{v_1 }}} \right)^{2L - 1 - k} } \right] \hfill \\
\end{gathered}
$}\\
\hline
&\multicolumn{2}{l}{}\\
$ \begin{gathered} \Pr\{D<0\} \\ (L = 1)\hfill \end{gathered} $& \multicolumn{2}{c}{$Q_1 (a,b)
- \dfrac{{v_2 /v_1 }} {{1 + v_2 /v_1 }}I_0
(ab)\exp \left\{ { - \frac{1} {2}(a^2  + b^2 )} \right\}$}\\
\hline
\end{tabular}
}
\end{center}
\label{table:1}

%% file: table2.tex
 \caption{Alternative expressions of the error rate for binary signals}
\begin{center}
{ \resizebox{!}{4cm}{
\begin{tabular}{c|c}
\sc{Parameter} & \sc{Definition} \\
 \hline\hline
&\\
$\delta_1 $  &  positive eigenvalue of $\mathbf{RQ}$ \\
\hline
&\\
$\delta_2 $  & negative eigenvalue of $\mathbf{RQ}$ \\
\hline
&\\
$a$  &  $ \sqrt {\dfrac{{2\delta_2 \left( {\sum\nolimits_{k = 1}^L {{\mathbf{m}}_k^H \left[
{\mathbf{Q}} - \delta_1 {\mathbf{R}^{ - 1} } \right]{\mathbf{m}}_k } } \right)}} {{\left(
{\delta_1  - \delta_2 } \right)^2 }}}$ \\
\hline
&\\
$b$  &  $ \sqrt {\dfrac{{2\delta_1 \left( {\sum\nolimits_{k = 1}^L {{\mathbf{m}}_k^H \left[
{\mathbf{Q}} - \delta_2 {\mathbf{R}^{ - 1} } \right]{\mathbf{m}}_k } } \right)}} {{\left(
{\delta_1  - \delta_2 } \right)^2 }}}$ \\
\hline \hline
&\\
$ \Pr\{D<0\} $ & $ \begin{gathered}
  Q_1 (a,b) - I_0 (ab)\exp \left\{ { - \frac{1}
{2}(a^2  + b^2 )} \right\} + \frac{{I_0 (ab)\exp \left\{ { - \frac{1} {2}(a^2  + b^2 )}
\right\}}} {{(1 - \delta_1 /\delta_2 )^{2L - 1} }}\sum\limits_{k = 0}^{L - 1} {\left(
{\begin{array}{*{20}c}
   {2L - 1}  \\
   k  \\
 \end{array} } \right)\left| {\frac{{\delta_1 }}
{{\delta_2 }}} \right|^k }  \hfill \\
   + \frac{{\exp \left\{ { - \frac{1}
{2}(a^2  + b^2 )} \right\}}} {{(1 - \delta_1 /\delta_2 )^{2L - 1} }}\sum\limits_{n = 1}^{L -
1} {I_n (ab)} \sum\limits_{k = 0}^{L - 1 - n} {\left( {\begin{array}{*{20}c}
   {2L - 1}  \\
   k  \\
 \end{array} } \right)}
    \left[ {\left( {\frac{b}
{a}} \right)^n \left| {\frac{{\delta_1 }} {{\delta_2 }}} \right|^k  - \left( {\frac{a} {b}}
\right)^n \left| {\frac{{\delta_1 }}
{{\delta_2 }}} \right|^{2L - 1 - k} } \right] \hfill \\
\end{gathered}$ \\
&\\
\hline $ \begin{gathered} \Pr\{D<0\} \\ (L = 1)\hfill \end{gathered}$ &  $Q_1 (a,b) +
\dfrac{{\delta_1 }} {{\delta_2 - \delta_1 }}I_0 (ab)\exp \left\{ { - \frac{1}
{2}(a^2  + b^2 )} \right\} $   \\
\hline
\end{tabular}
} }
\end{center}
\label{table:2}

%% file: table3.tex
\begin{table}[ht!]
\caption{Parameters calculation}
\begin{center}
{
\begin{tabular}{c|c|c}
\sc{Parameter} & \sc{Proakis} & \sc{Corrected} \\
 \hline\hline
$w$            & 0                   &  0 \\
$v_1$          & 1                   &  1 \\
$v_2$          & 1                   &  1 \\
$\alpha_{11}$  & 2                   &  2 \\
$\alpha_{21}$  & 2                   &  0 \\
$a$            & 0                   &  1 \\
$b$            & $\sqrt{2}$          & 1 \\
$\Pr\{D<0\}$    & 0.18394            & 1/2 \\
\end{tabular}
}
\end{center}
\label{table:3}
\end{table}

%% file: CommentsProakisAnalysisQuadraticforms.bbl
\begin{thebibliography}{1}

\bibitem{Proakis68}
J. G. Proakis, ``On the probability of error for multichannel reception of binary signals''
{\em IEEE Trans. Commun. Technol.}, vol. 16, pp. 68-71, Feb 1968.

\bibitem{Proakis01}
J. G. Proakis, {\em Digital Communications}, 4th ed., New York, Mc
Graw-Hill, 2001.

\bibitem{Simon05}
M.K. Simon, M. Alouini, {\em Digital Communication over Fading channels}, 2nd ed., New Jersey,
John Wiley \& Sons, 2005.

\bibitem{Simon98}
M.K. Simon, M. Alouini, ``A Unified Approach to the Performance Analysis of Digital
Communication over Generalized Fading Channels'' in {\em Proc. IEEE}, vol. 86, no. 9, pp.
1860-1877, Sep. 1998

\bibitem{Sudhanshu03}
S. Gaur, A. Annamalai, ``Some Integrals Involving the $Qm(a \sqrt x,b \sqrt x)$ With
Application to Error Probability Analysis of Diversity Receivers'' {\em IEEE Trans. on
Vehicular Technology}, vol. 52, no. 6, pp. 1568-1575, Nov 2003.

\bibitem{Cao05}
L. Cao, C. Beaulieu,``Closed-Form BER Results for MRC Diversity With Channel Estimation Errors
in Ricean Fading Channels'' {\em IEEE Trans. on Wireless Communications}, vol. 4, no. 4, pp.
1440-1447, July 2005.

\bibitem{Jun00}
J. Lu, T. T. Tjhung, F. Adachi, C.L. Huang, ``BER Performance of OFDM-MDPSK System in
Frequency-Selective Rician Fading with Diversity Reception'' {\em IEEE Trans. on Vehicular
Technology}, vol. 49, no. 4, pp. 1216-1225, Jul. 2000.


\bibitem{McCloud02}
M.L. McCloud, M.K. Varanasi, ``Modulation and Coding for Noncoherent Communications'' {\em
Journal of VLSI Signal Processing}, no. 30, pp. 35-54, 2002.

\bibitem{Biyari93}
K. H. Biyari, W.C. Lindsey, ``Statistical Distribution of Hermitian Quadratic Forms in Complex
Gaussian Variables'' {\em IEEE Trans. on Information Theory}, vol. 39, no. 3, pp. 1076-1082,
May 1993.

\bibitem{turin97}
G.L. Turin, ``The characteristic function of Hermitian quadratic forms in complex normal
variables'' {\em Biometrika}, vol. 47, no. 1-2, pp. 199--201, 1960.

\bibitem{lindsey64}
W.C. Lindsey, ``Error Probabilities for Rician Fading Multichannel Reception of Binary and
N-ary Signals'' {\em IEEE Trans. on Information Theory}, vol. IT-10, pp. 339-350,Oct. 1964.

\end{thebibliography}
